\documentclass[modern]{aastex7}

\newcommand{\ecss}{erg~cm$^{-2}$~s$^{-1}$~sr$^{-1}$}

\newcommand{\kms}{km~s$^{-1}$}
\newcommand{\as}{$^{\prime\prime}$}
\renewcommand{\ion}[2]{#1\,\textsc{#2}}

\submitjournal{ApJ}

\graphicspath{{./}{figures/}}

\begin{document}

\title{Fe\,XVIII and Fe\,XX Forbidden Lines Observed by Solar Orbiter/SPICE}

\author[0000-0001-9034-2925]{Peter R. Young}
\affiliation{NASA Goddard Space Flight Center, Heliophysics Science Division, Code 671, 8800 Greenbelt Rd., Greenbelt, MD 20771, USA}
\affiliation{Northumbria University, Newcastle upon Tyne, NE1 8ST, UK}
\email{peter.r.young@nasa.gov}

\author[0000-0003-0656-2437]{Andrew R. Inglis}
\affiliation{NASA Goddard Space Flight Center, Heliophysics Science Division, Code 671, 8800 Greenbelt Rd., Greenbelt, MD 20771, USA} 
\affiliation{Department of Physics, Catholic University of America, 620 Michigan Avenue, Northeast, Washington, DC 20064, USA}
\email{andrew.inglis@nasa.gov}

\author[0000-0001-5316-914X]{Graham S. Kerr}
\affiliation{NASA Goddard Space Flight Center, Heliophysics Science Division, Code 671, 8800 Greenbelt Rd., Greenbelt, MD 20771, USA} 
\affiliation{Department of Physics, Catholic University of America, 620 Michigan Avenue, Northeast, Washington, DC 20064, USA}
\email{graham.s.kerr@nasa.gov}

\author[0000-0001-9632-447X]{Therese A. Kucera}
\affiliation{NASA Goddard Space Flight Center, Heliophysics Science Division, Code 671, 8800 Greenbelt Rd., Greenbelt, MD 20771, USA}
\email{therese.a.kucera@nasa.gov}

\author[0000-0001-8661-3825]{Daniel F. Ryan}
\affiliation{University College London, Mullard Space Science Laboratory, Holmbury St Mary, Dorking, Surrey RH5 6NT, UK}
\email{daniel.ryan@ucl.ac.uk}

\begin{abstract}
The first simultaneous observations of the \ion{Fe}{xviii} 974.86~\AA\ and \ion{Fe}{xx} 721.56~\AA\ forbidden lines from the Spectral Imaging of the Coronal Environment (SPICE) spectrograph on Solar Orbiter are presented. The lines were observed from the post-flare loops of an M2.5 class solar flare that peaked at 23:49~UT on 2024 March 23. 
The \ion{Fe}{xx}/\ion{Fe}{xviii} ratio is used to derive a temperature space-time map for the flaring period, with values ranging from 8 to 20~MK. The map reveals repeated episodes of heating at the SPICE slit location over a 30~min period. For one location with the brightest emission, the plasma cooled from 10~MK to 8~MK in 5~min which is longer than the expected conductive cooling time of 170~s, suggesting continued background heating during the cooling period. Doppler shifts of between 0 and $+10$~\kms\ were obtained with precisions of 1--4~\kms, but the accuracies are lower due to uncertainties over the absolute wavelength calibration hence we can not conclude there are plasma flows in the flare loops.
The widths of the two lines were found to be close to the instrumental widths with no evidence of non-thermal broadening, although this result is limited by the instrument resolution.
The \ion{Fe}{xviii} and \ion{Fe}{xx} lines have high signal-to-noise with only a 5~s exposure time, demonstrating that the lines will be valuable for high-cadence flare studies with SPICE.
\end{abstract}

\section{Introduction}

Solar and stellar flares produce plasma with temperatures in the range 6 to 30~MK, giving copious emission at soft X-rays principally due to hydrogen and helium-like ions of elements oxygen through iron, and the iron sequence \ion{Fe}{xvii--xxiii}. The latter ions also yield emission lines in the far ultraviolet region from 500 to 1400~\AA\ through so-called forbidden transitions. These are transitions that occur between levels of the ions' ground configurations and have much smaller radiative decay rates than the X-ray lines. 
Despite being ``forbidden," the photon yields from the UV lines of \ion{Fe}{xviii--xxiii} are comparable to the strongest resonance lines of the ions at soft X-ray wavelengths. This is demonstrated in Table~\ref{tbl.lines} where theoretical line intensities for the strongest X-ray and strongest UV line for each of the six ions are given. The intensities are calculated from CHIANTI 11.0 \citep{2024ApJ...974...71D} using the flare differential emission measure distributed with CHIANTI \citep{1979ApJ...229..772D}. The strongest UV forbidden lines of \ion{Fe}{xviii} and \ion{Fe}{xxi} are stronger than all of the ions' X-ray lines, while for \ion{Fe}{xix}, \ion{Fe}{xx} and \ion{Fe}{xxii} the UV lines are within a factor two of the strongest X-ray lines. The \ion{Fe}{xviii} and \ion{Fe}{xx} UV lines are observed by SPICE and studied in the present work (indicated by bold font in Table~\ref{tbl.lines}). 

All of the UV lines in Table~\ref{tbl.lines} were observed by the Solar Ultraviolet Measurements of Emitted Radiation \citep{1995SoPh..162..189W} experiment on board the Solar and Heliospheric Observatory (SOHO). For example, \citet{2000ApJ...544..508F} and \citet{2003ApJ...582..506L} measured  all of the UV lines from Table~\ref{tbl.lines} from a flare spectrum obtained on 1999 May 9. However, SUMER was restricted to observing flares and active regions at the solar limb due to count rate concerns for on-disk active regions. In addition, SUMER could only observe a narrow 43~\AA\ region in a single exposure thus it was not possible to observe two or more of the Table~\ref{tbl.lines} lines simultaneously, although combinations of the Table~\ref{tbl.lines} lines with weaker forbidden lines could be observed simultaneously. For example, \ion{Fe}{xx} 821.79~\AA\ and \ion{Fe}{xxi} 845.57~\AA, which were used to derive plasma temperature in  \citet{2003ApJ...582..506L}.

The \ion{Fe}{xxi} 1354.06~\AA\ line is observed by the Interface Region Imaging Spectrograph \citep[IRIS:][]{2014SoPh..289.2733D}, and was first reported by \citet{2015ApJ...799..218Y} from a flare on 2014 March 29. The high spectral and spatial resolutions of IRIS enable Doppler shifts and non-thermal broadening to be measured \citep[e.g., see also][]{2015ApJ...807L..22G,2019ApJ...879L..17P}, and allowed the location of  \ion{Fe}{xxi} emission to be accurately placed relative to chromospheric emission in flare ribbons. One complication is the presence of a number of chromospheric lines in flare ribbon spectra that can hamper line fitting, which is not the case for the SPICE lines.

Forbidden lines longward of the hydrogen Lyman edge at 912~\AA\ also provide valuable diagnostics of stellar coronae. \ion{Fe}{xxi} 1354.06~\AA\ has been measured in spectra from the Hubble Space Telescope's Goddard High Resolution Spectrograph \citep{1994ApJ...421..800M} and Space Telescope Imaging Spectrograph \citep{2002ApJ...565L..97J}. \ion{Fe}{xviii} 974.86~\AA\ and \ion{Fe}{xix} 1118.06~\AA\ were present in spectra from the Far Ultraviolet Spectroscopic Explorer (FUSE), with the former measured in Capella \citep{2001ApJ...555L.121Y} and a survey of both lines across a range of stars was performed by \citet{2003ApJ...585..993R}. Generally, the UV lines show little non-thermal broadening and no Doppler shifts in the stellar spectra, suggesting hot plasma contained in stable magnetic structures rather than the dynamic plasma found in solar flares.

In the present article, SPICE data from a solar flare observed on 2024 March 23 that displayed strong \ion{Fe}{xviii} and \ion{Fe}{xx} emission are presented. Details of the observation are given in Section~\ref{sec:obs}, and Section~\ref{sec:analysis} describes how the lines were measured and physical parameters derived. Section~\ref{sec:summary} summarizes the results, and discusses how SPICE observations can be optimized for observing hot emission in flares.

\begin{deluxetable}{lccccc}[t]
\tablecaption{A comparison of intensities for the strongest X-ray and UV lines of \ion{Fe}{xviii--xxiii}. Bold font indicates the lines observed by SPICE.\label{tbl.lines}}
\tablehead{
  &\multicolumn{2}{c}{X-ray lines} &
  &\multicolumn{2}{c}{UV lines} \\
  \cline{2-3}\cline{5-6}
  \colhead{Ion} &
  \colhead{Wavelength} &
  \colhead{Intensity\tablenotemark{a}} &&
  \colhead{Wavelength} &
  \colhead{Intensity} \\
  &\colhead{(\AA)} & & & \colhead{(\AA)} &
}
\startdata
\ion{Fe}{xviii} & 93.93 & 19 && \textbf{974.86} & 22 \\
\ion{Fe}{xix} & 108.36 & 25 && 1118.06 & 24 \\
\ion{Fe}{xx} & 132.84 & 35 && \textbf{721.56} & 19 \\
\ion{Fe}{xxi} & 128.75 & 59 && 1354.06 & 73 \\
\ion{Fe}{xxii} & 117.14 & 46 && 845.55 & 35 \\
\ion{Fe}{xxiii} & 132.91 & 121 && 1079.41 & 4 \\
\noalign{\smallskip}
\enddata
\tablenotetext{a}{Units: $10^{14}$~photon\,cm$^{-2}$~s$^{-1}$~sr$^{-1}$.}
\end{deluxetable}

\section{Observations}\label{sec:obs}

\begin{figure}[t]
    \centering
    \includegraphics[width=\linewidth]{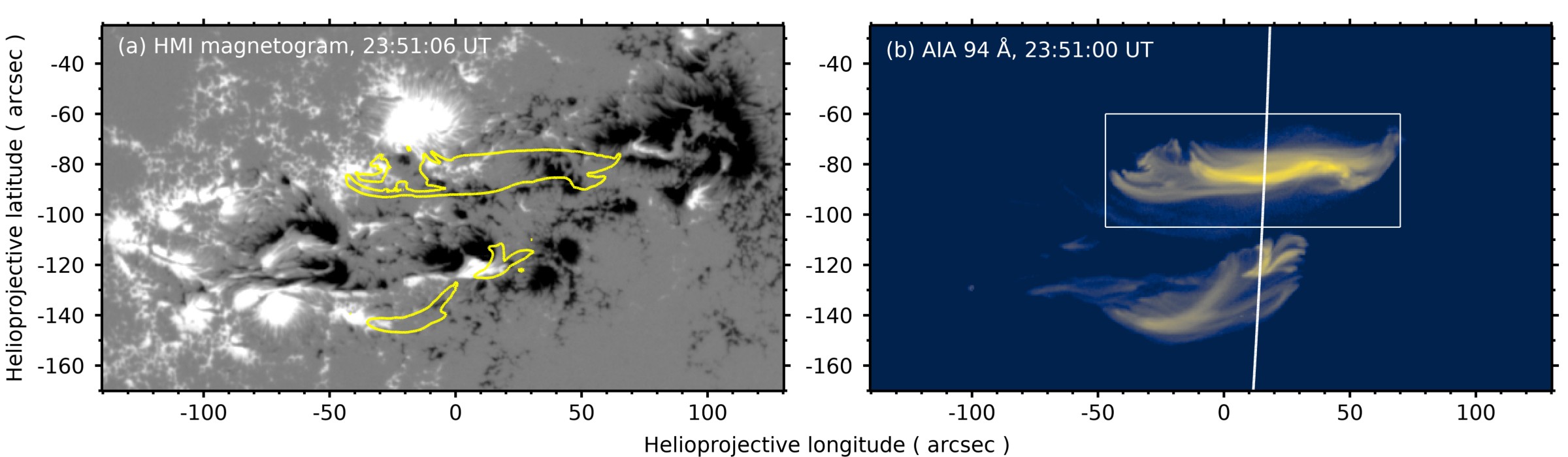}
    \caption{Images of active region AR 13615. Panel (a) shows an HMI LOS magnetogram  saturated at $\pm$ 500~G. Panel (b) shows a co-temporal AIA 94~\AA\ image displayed with a logarithmic intensity scaling. The 94~\AA\ image is over-plotted on Panel (a) as a yellow contour corresponding to 300~DN~pix$^{-1}$~s$^{-1}$. The white line on Panel (b) shows the position of the SPICE slit, and the white box shows the sub-region displayed in Figure~\ref{fig:movie}.}
    \label{fig:context}
\end{figure}

Active region AR 13615 crossed the solar disk, as viewed from Earth, during March 16--30 with central meridian passage occurring around 0~UT on March 24. The region was very active and was selected as the target for the first Solar Orbiter Major Flare campaign\footnote{SOOP name: L\_BOTH\_HRES\_HCAD\_Major-Flare} that yielded several observations of AR 13615 during March 19 to April 6 (Ryan et al., submitted). The observations presented here are from March 23, and Figure~\ref{fig:context} shows a Helioseismic and Magnetic Imager  \citep[HMI:][]{2012SoPh..275..207S} line-of-sight (LOS) magnetogram and an AIA 94~\AA\ image from this day at 23:51~UT. The latter image is dominated by emission from \ion{Fe}{xviii} 93.93~\AA\ (Table~\ref{tbl.lines}), formed at 7~MK.

\begin{figure}[t]
    \centering
    \begin{interactive}{animation}{movie.mp4}
    \includegraphics[width=0.7\linewidth]{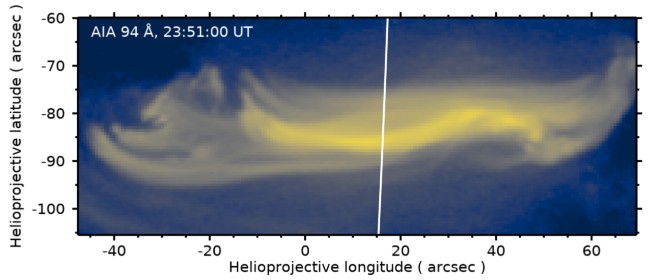}
    \end{interactive}
    \caption{An AIA 94~\AA\ image of the flare loops studied in the present work. A logarithmic intensity scaling  is used. The white line indicates the location of the SPICE slit. An animated version of this figure is available, showing the evolution of the loops from 23:30~UT on 2024 March 23 to 00:30~UT on March 24.}
    \label{fig:movie}
\end{figure}

Figure~\ref{fig:context}(b) shows two sets of hot loops running approximately east-west at locations $y=-80$ and $y=-130$. These loop sets can be matched to two  bipolar regions in the  HMI magnetogram (Figure~\ref{fig:context}(a)). The lower loop set produced three M-class flares during 14--17~UT with GOES classifications of M2 and higher\footnote{\href{Hinode flare catalog}{https://hinode.isee.nagoya-u.ac.jp/flare\_catalogue/} \citep{2012SoPh..279..317W}.}. An M2.5 class flare then occurred in the upper region, peaking at 23:49~UT, and this is the event studied in the present work. None of the AR 13615 flares produced eruptions during this time period. An animation showing the evolution of the flare loops in AIA 94~\AA\ during the period 23:30~UT to 00:30~UT is shown in Figure~\ref{fig:movie}. There is a visual impression of the loop bundle beginning twisted in this sequence and becoming straighter by the end of the sequence (00:06~UT onward), suggesting a release of magnetic energy.

The second of the Solar Orbiter Major Flare campaign's observing windows was active between March 23 21:55~UT and March 24 03:00~UT, during which period Solar Orbiter was at a distance of 0.38 AU, and 11$^\circ$ ahead of the Earth in longitude.
The difference in time between when photons arrived at Solar Orbiter and when they arrived at Earth is 307~s. Thus
a feature observed by Solar Orbiter at 23:50:00~UT, for example, would be observed at 23:55:07~UT by an Earth-orbiting spacecraft. All times quoted in the present article will be Earth times, unless otherwise stated.

We note that both the EUV Imaging Spectrometer \citep[EIS:][]{2007SoPh..243...19C} on the Hinode spacecraft  and IRIS supported the SOOP, but neither dataset are used here. EIS was in flare trigger mode, waiting for the flare flag from the X-Ray Telescope (XRT). However, the flag was not raised despite XRT observing the flare. IRIS was in sit-and-stare mode and the slit was around 15\as\ to the west of the SPICE slit. It thus observed the post-flare loops, but the  short exposure time used (0.3~s) was too short for studying the \ion{Fe}{xxi} 1354.1~\AA\ line, which would have complemented the SPICE data.

\subsection{SPICE Observations}

SPICE's role within the SOOP was to provide high-cadence EUV flare observations. Thus, it was operated in a sit-and-stare mode with 35 identical rasters beginning at 22:30~UT and completing at 01:47~UT (SPICE times). Solar feature tracking was on for the observation, and each raster consisted of 64 exposures with exposure times of 4.8~s, and the 4\as\ slit was used. The cadence of the exposures was 5.1~s giving each raster a duration of 5~min 26~s. Nine wavelength windows were downloaded, and windows 2 and 5 contain the \ion{Fe}{xx} and \ion{Fe}{xviii} lines, respectively. For both windows, $\times$2 spectral binning was performed and 40 binned spectral pixels were downloaded, corresponding to wavelength regions of width 7.80~\AA, and 7.70~\AA, respectively. Each window contains additional lines to the iron lines, and blending issues are discussed in Section~\ref{sec:blend}.

In addition to the March 23 dataset, we also utilized two spectral atlas datasets obtained on 2024 March 24 and 2024 October 1 that obtained the complete spectral range of SPICE. The first was a raster during 03:07--04:00~UT consisting of 30 exposures with the 2\as\ slit, each of 60~s, that observed active region fan loops from AR 13615 and is used in Section~\ref{sec:blend}. The second was a 10-step raster with the 4\as\ slit and 60~s exposure times that was run during 00:37--00:55~UT. The target was active region AR 13842, and the data are used in Section~\ref{sec:profile}.

The SPICE data used in the current work are from SPICE Data Release 5.0 \citep{https://doi.org/10.48326/idoc.medoc.spice.5.0}, and the analysis software routines are from the IDL \textit{Solarsoft} distribution \citep{1998SoPh..182..497F,2012ascl.soft08013F}.

\subsection{Coalignment}

The SPICE dataset was coaligned with AIA by making use of the World Coordinate System (WCS) software available in \textit{Solarsoft}. This enables the SPICE slit to be over-plotted on a co-temporal AIA image. It was found necessary to manually adjust the position of the SPICE slit in order to match spatial features in AIA data. For this purpose, 
an intense flare ribbon brightening observed by SPICE at 23:48:35~UT in chromospheric and transition region lines was coaligned with the 1600~\AA\ channel of AIA. A correction of $(-8$\as,$-18$\as) to the SPICE slit position in the AIA image plane was determined. This offset has been applied to the slit locations shown in Figures~\ref{fig:context} and \ref{fig:movie}.

\section{Data analysis}\label{sec:analysis}

The procedures for fitting the SPICE emission lines and deriving physical parameters are described here. Atomic data, including wavelength information, are obtained from version 11 \citep{2024ApJ...974...71D} of the CHIANTI atomic database \citep{1997A&AS..125..149D,2016JPhB...49g4009Y}.

\subsection{Wavelength Information and Blending Issues}\label{sec:blend}

\ion{Fe}{xviii} 974.86~\AA\ corresponds to the $2p^5$ $^2P_{3/2}$--$2p^5$ $^2P_{1/2}$ transition and was first reported by \citet{1975ApJ...196L..83D}. The reference wavelength of $974.86\pm 0.02$~\AA\ is obtained from Tokamak spectra \citep{1984PhST....8...10P} and the Ritz wavelength in CHIANTI is 974.858~\AA, which is estimated to be accurate to 0.01~\AA\ \citep{2006A&A...459..307D}.
Studies of the \ion{Fe}{xviii} line from FUSE stellar spectra \citep{2003ApJ...585..993R} and SOHO/SUMER spectra \citep{2012ApJ...754L..40T} suggest the line is not affected by blending in active region conditions. However, the line lies between the strong \ion{H}{i} 972.54~\AA\ (Ly-$\gamma$) and \ion{C}{iii} 977.02~\AA\ lines that impact measurements of the \ion{Fe}{xviii} line because of the extended line wings discussed in Section~\ref{sec:profile}. 

\citet{2000ApJ...538..424K} measured the \ion{Fe}{xx} $2s^22p^3$ $^4S_{3/2}$--$2s^22p^3$ $^4D_{3/2}$ transition for the first time from spectra obtained with SOHO/SUMER, giving a wavelength of $721.55\pm 0.02$~\AA. By combining with measurements of the remaining forbidden lines, a Ritz wavelength of $721.558\pm 0.010$~\AA\ was obtained, which is the value found in CHIANTI.
The  line is partly blended with the \ion{Fe}{viii} $3p^64p$ $^2P_{3/2}$--$3p^64d$ $^2D_{5/2}$ transition that was first identified by  \citet{2003ApJ...595..517E}. The reference wavelength from laboratory spectra is $721.268\pm 0.010$~\AA, which is consistent with the wavelength of $721.26\pm 0.03$~\AA\ obtained from off-limb SUMER spectra by \citet{2004A&A...427.1045C}.

\begin{figure}[t]
    \centering
    \includegraphics[width=0.9\linewidth]{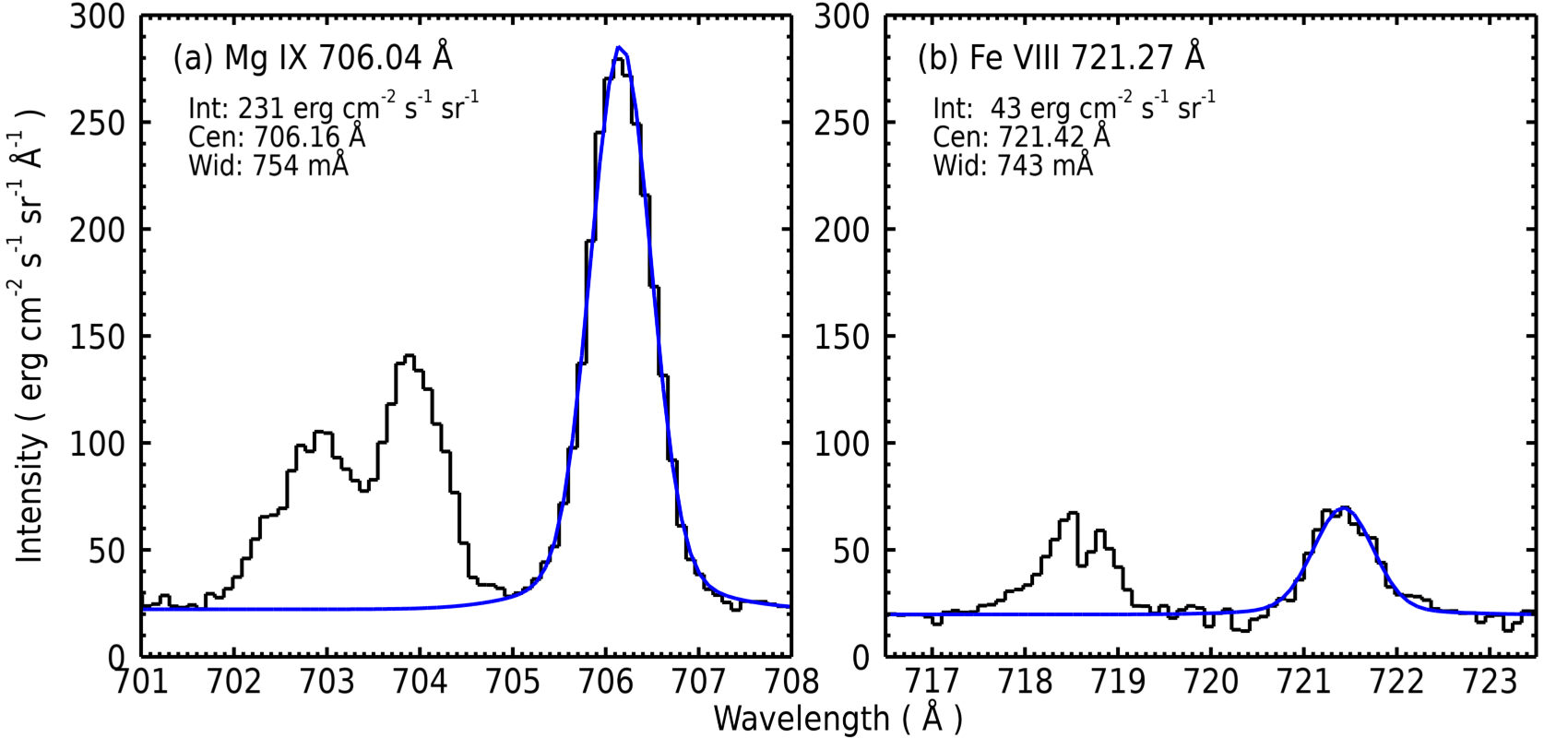}
    \caption{Spectra from an active region fan loop observed on 2024 March 24 showing the (a) \ion{Mg}{ix} 706.04~\AA\ and (b) \ion{Fe}{viii} 721.27~\AA\ lines. The blue lines show fits to the lines using the two-Gaussian profiles, and the derived intensity (int), centroid (cen) and FWHM (wid) are displayed on each panel. The line groups at 702--705~\AA\ and 718--719~\AA\ are due to \ion{O}{iii} and \ion{O}{ii}, respectively.}
    \label{fig:fe8}
\end{figure}

\ion{Fe}{viii} 721.27~\AA\ is generally expected to be weak: CHIANTI quiet Sun and active region models put the \ion{Mg}{ix} 706.04~\AA/\ion{Fe}{viii} 721.27~\AA\ ratio to be around 7--8 and the \ion{Mg}{ix} line itself is rather weak \citep[see, e.g., Figure~1 of][]{2021A&A...656A..38F}. \ion{Fe}{viii} lines are enhanced in active region fan loops \citep{2007PASJ...59S.727Y}, and the fan loop spectral atlas from March 24 is shown in Figure~\ref{fig:fe8}, where the \ion{Mg}{ix} and \ion{Fe}{viii} emission lines are well observed. The intensity ratio is 5.3, and the line widths are very similar. 
Using a rest wavelength of $706.04\pm 0.03$~\AA\ for the \ion{Mg}{ix} line from \citet{2004A&A...427.1045C} and the laboratory reference wavelength for the \ion{Fe}{viii} line quoted above, the Doppler shifts are 51~\kms\ and 66~\kms\ for \ion{Mg}{ix} and \ion{Fe}{viii}, respectively. These values are consistent within the fitting and reference wavelength uncertainties, giving confidence that the \ion{Fe}{viii} identification is accurate. The derived velocities are rather large for fan loops, but no attempt has been made to perform an absolute wavelength calibration for this dataset as the absolute velocities are not relevant for the present analysis.

In summary, the \ion{Mg}{ix} 706.04~\AA\ line is valuable in estimating the strength of the \ion{Fe}{viii} 721.27~\AA\ line if there are concerns that it is making a significant contribution to the \ion{Fe}{xx} 721.56~\AA\ line. Dividing the \ion{Mg}{ix} intensity by a factor five should yield an upper limit to the \ion{Fe}{viii} intensity. For the post-flare loop spectra studied here the \ion{Fe}{xx} line is at least a factor 100 above this level at most locations, and so \ion{Fe}{viii} can be safely ignored.

\subsection{Line Profile Models}\label{sec:profile}

The SPICE emission lines exhibit two features that inhibit the derivation of Doppler velocities and line widths. The first is a tilt of the point spread function (PSF) in wavelength--$y$ space as viewed on the detector, which has the consequence of introducing an artificial Doppler shift to the line at locations where there is an intensity gradient in the $y$ direction (along the slit).
The effect is most readily seen at the limb and in compact bright points. The second feature is the presence of enhanced emission in the line wings, above what a single Gaussian fit can reproduce. Both effects were discussed in \citet{2023A&A...678A..52P}, where a method for deconvolving the PSF from the data is presented. The prescription for applying the method requires adjusting parameters for the particular dataset or emission line under study. For the present work, a simpler approach is taken whereby the emission line wings are modeled by fitting an additional, broadened Gaussian. In addition, to avoid artifacts due to the tilted PSF, spectra are spatially-averaged in the $y$-direction across compact structures in order that the Doppler effects are canceled out (Section~\ref{sec:gauss}).

The emission lines are modeled with a narrow Gaussian for the core of the line, and a broad Gaussian for the wings. The latter is forced to have the same centroid as the narrow Gaussian. If the peak of the narrow Gaussian is $P$ and the full-width at half-maximum (FWHM) is $W$, then the peak and FWHM of the broad Gaussian are set to $aP$ and $bW$. For lines in the SW channel, $a=0.062$ and $b=2.55$, and for lines in the LW channel, $a=0.070$ and $b=2.27$. These parameters were obtained by performing two Gaussian fits to \ion{Fe}{xx} 721.56~\AA\ and \ion{O}{vi} 1031.91~\AA\ in the active region spectral atlas dataset from 2024 October 1. For these fits, the amplitudes and widths of the broad Gaussian were free to vary. Both lines are strong and relatively isolated in this dataset, enabling high-quality fits. The fits at different spatial pixels were visually inspected and fits from  single pixels that were judged to be representative were used to define the $a$ and $b$ parameters.

\subsection{Temperature Map}\label{sec:temp}

\begin{figure}[t]
    \centering
    \includegraphics[width=1.0\linewidth]{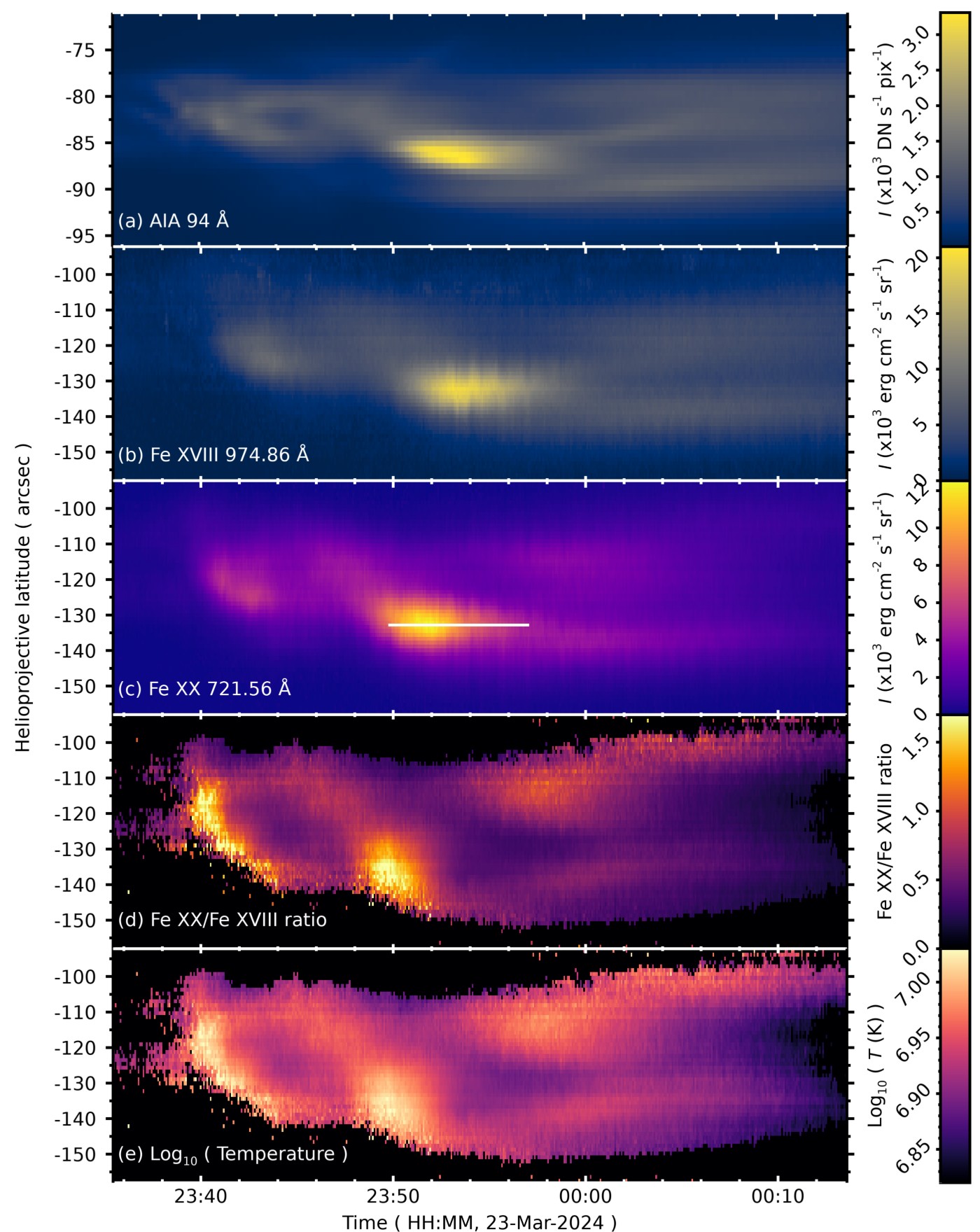}
    \caption{Panel (a) shows an image of AIA 94~\AA\ intensity, $I$,  versus time for a slit corresponding to the location of the SPICE slit. Panels (b) and (c) show intensity maps obtained from Gaussian-fitting of \ion{Fe}{xviii} 974.86~\AA\ and \ion{Fe}{xx} 721.56~\AA, respectively. Panel (d) shows the \ion{Fe}{xx}/\ion{Fe}{xviii} intensity ratio, and panel (e) shows the temperature map derived from the ratio map using CHIANTI. The white line on panel (c) shows the region analyzed in Section~\ref{sec:gauss}.}
    \label{fig:temp}
\end{figure}

\begin{figure}[t]
    \centering
    \includegraphics[width=0.7\linewidth]{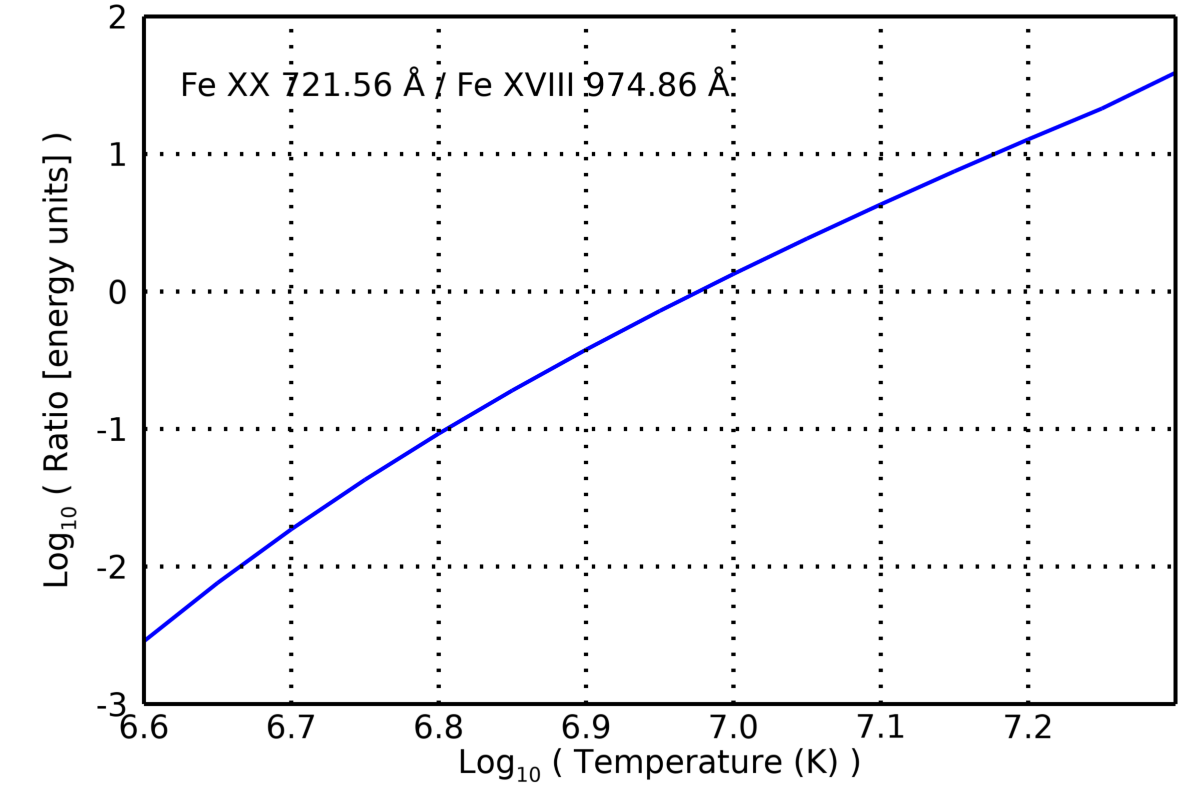}
    \caption{Predicted variation of the \ion{Fe}{xx} 721.56~\AA\ / \ion{Fe}{xviii} 974.86~\AA\ ratio as a function of temperature, derived using CHIANTI 11.0.}
    \label{fig:chianti}
\end{figure}

Accurate line width and Doppler velocity time series images along the SPICE slit are not possible due to the point spread function artifacts discussed in Section~\ref{sec:profile}. However, line intensity images can be created and are shown in Figure~\ref{fig:temp}. Data from seven consecutive SPICE rasters have been concatenated, corresponding to the time interval 23:35~UT to 00:13~UT. Panels (b) and (c) show the intensity images from the \ion{Fe}{xviii} and \ion{Fe}{xx} lines.
Both images show striations in the $y$ direction that are most clearly seen above and below the brightest features of the images. They correspond to shifts of about 1~pixel in the intensity profile along the slit caused by spacecraft jitter and they have a period of around 1~min.
Panel (d) shows the \ion{Fe}{xx}/\ion{Fe}{xviii} intensity ratio, which is converted to a temperature (Panel (e)) using the data shown in Figure~\ref{fig:chianti}. The latter is derived from CHIANTI under the assumption of ionization equilibrium at a density of $10^{10}$~cm$^{-3}$. The ratio is insensitive over the range of typical coronal densities ($10^8$--$10^{12}$~cm$^{-3}$).  Panel (a) shows a plot of AIA 94~\AA\ intensity versus time obtained by extracting a vertical slice through the AIA images at a location corresponding to the SPICE slit (white lines on Figures~\ref{fig:context} and \ref{fig:movie}). The intensity was averaged across five AIA pixels in the $x$ direction to allow for uncertainties in the SPICE slit position. The  AIA image shows good morphological agreement with the SPICE \ion{Fe}{xviii} image (Panel b), which is consistent with the AIA channel being dominated by emission from \ion{Fe}{xviii} 93.93~\AA\ (Table~\ref{tbl.lines}). The comparison also demonstrates the higher spatial resolution of AIA compared to SPICE even though SPICE is around a factor three closer to the Sun.

The $\sim$\,10~MK plasma produced during flares can be produced by direct heating of coronal plasma or by strong heating of chromospheric plasma that subsequently rises into the corona. The heated plasma then cools by thermal conduction and/or radiation losses \citep[e.g.,][]{2013ApJ...778...68R}. The temperature map (Figure~\ref{fig:temp}(e)) shows a complex pattern that is likely due to multiple heating events during the flare. For example, temperatures of 20~MK are found initially at 23:40~UT for locations $y=\mathbf{-130}$ to $-110$. Further heating occurs beginning at 23:43~UT, with the strongest emission found at $y=-135$ where the temperature again reaches 20~MK. A subsequent heating event takes places from 23:52~UT onward at $y=-120$ to $y=-100$, although the maximum temperature is lower at 9~MK. Note also that the temperature features show drifts in $y$ location with time, which is consistent with the apparent untwisting of loops seen in the AIA 94~\AA\ image sequence (Figure~\ref{fig:movie}).

\subsection{Gaussian Fitting and Results}\label{sec:gauss}

\begin{figure}[t]
    \centering
    \includegraphics[width=0.9\linewidth]{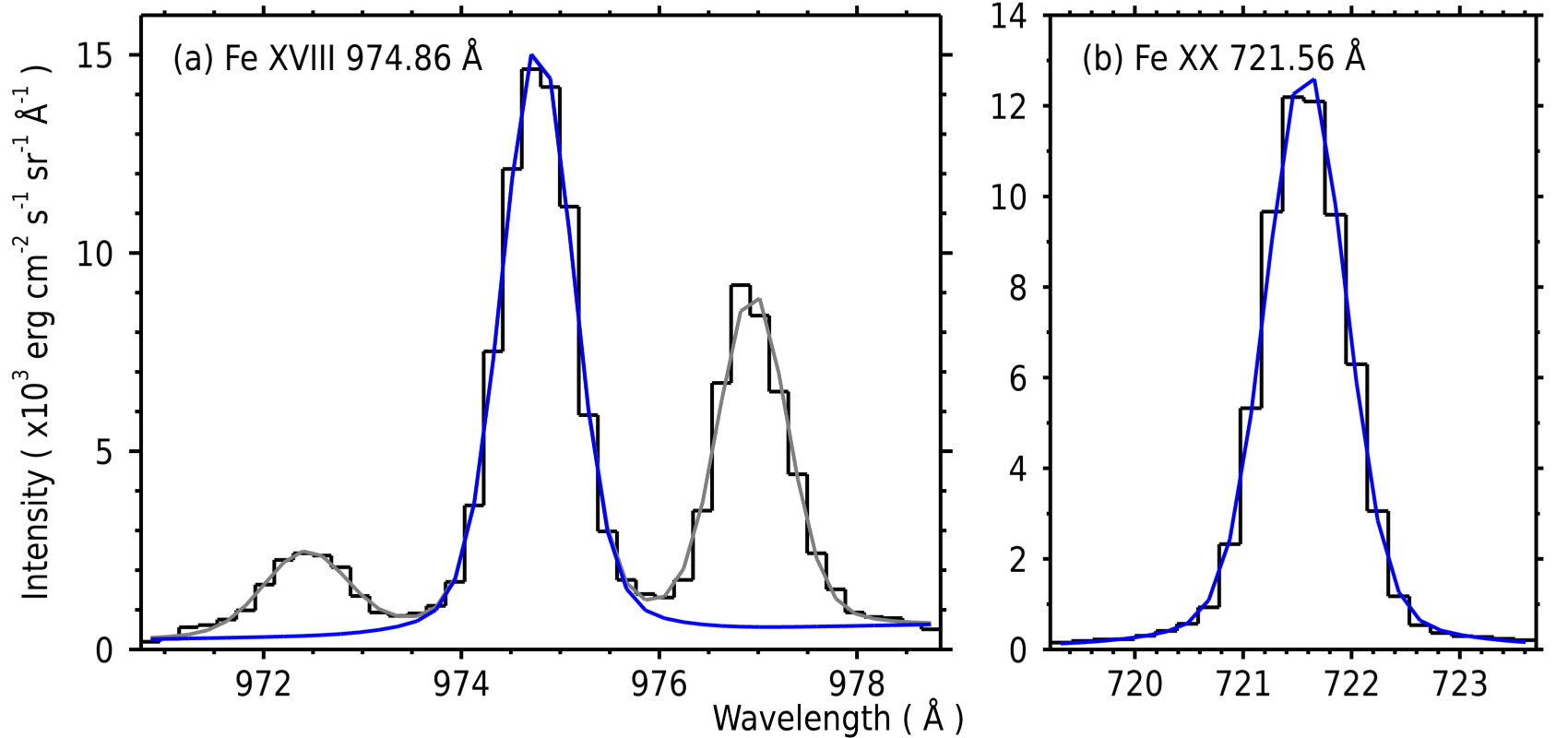}
    \caption{Spectra from the post-flare loops at 23:46:32~UT, showing (a) the \ion{Fe}{xviii} 974.86~\AA\  line, and (b) the \ion{Fe}{xx} 721.56~\AA\ line. The fits to the lines are over-plotted in blue. The gray lines in panel (a) show the total fit function that includes \ion{H}{i} 972.54~\AA\  and \ion{C}{iii} 977.02~\AA.}
    \label{fig:fits}
\end{figure}

The previous section demonstrated how temperature maps can be derived from the ratio of the \ion{Fe}{xviii} and \ion{Fe}{xx} line intensities. In this section, line widths and Doppler velocities are derived, but only for a single $y$-position that corresponds to a ridge of peak intensity in the two lines. This is necessary to avoid the PSF artifacts described in Section~\ref{sec:profile}. The region selected is indicated by the horizontal white line in Figure~\ref{fig:temp}(b). For each exposure along this line, 
the $y$-pixel with the peak intensity was found and the spectra were averaged over five $y$-pixels centered on this pixel using the IDL routine \textsf{spice\_mask\_spectrum}. Fit templates for each wavelength window were then created using the EIS software routine \textsf{eis\_fit\_template}. For \ion{Fe}{xx} the wavelength range was restricted so that only the hot line was included, excluding the \ion{O}{ii} lines (Figure~\ref{fig:fe8}(b)). For \ion{Fe}{xviii} the template included all three emission lines in this window (Section~\ref{sec:blend}). Spectral fitting was performed by inputting the mask spectra and templates to the routine \textsf{spice\_mask\_auto\_fit}. Example fits are shown in Figure~\ref{fig:fits}, which are taken from exposure 62 of the raster beginning 23:41~UT (SPICE time).

\begin{figure}[t]
    \centering
    \includegraphics[width=0.6\linewidth]{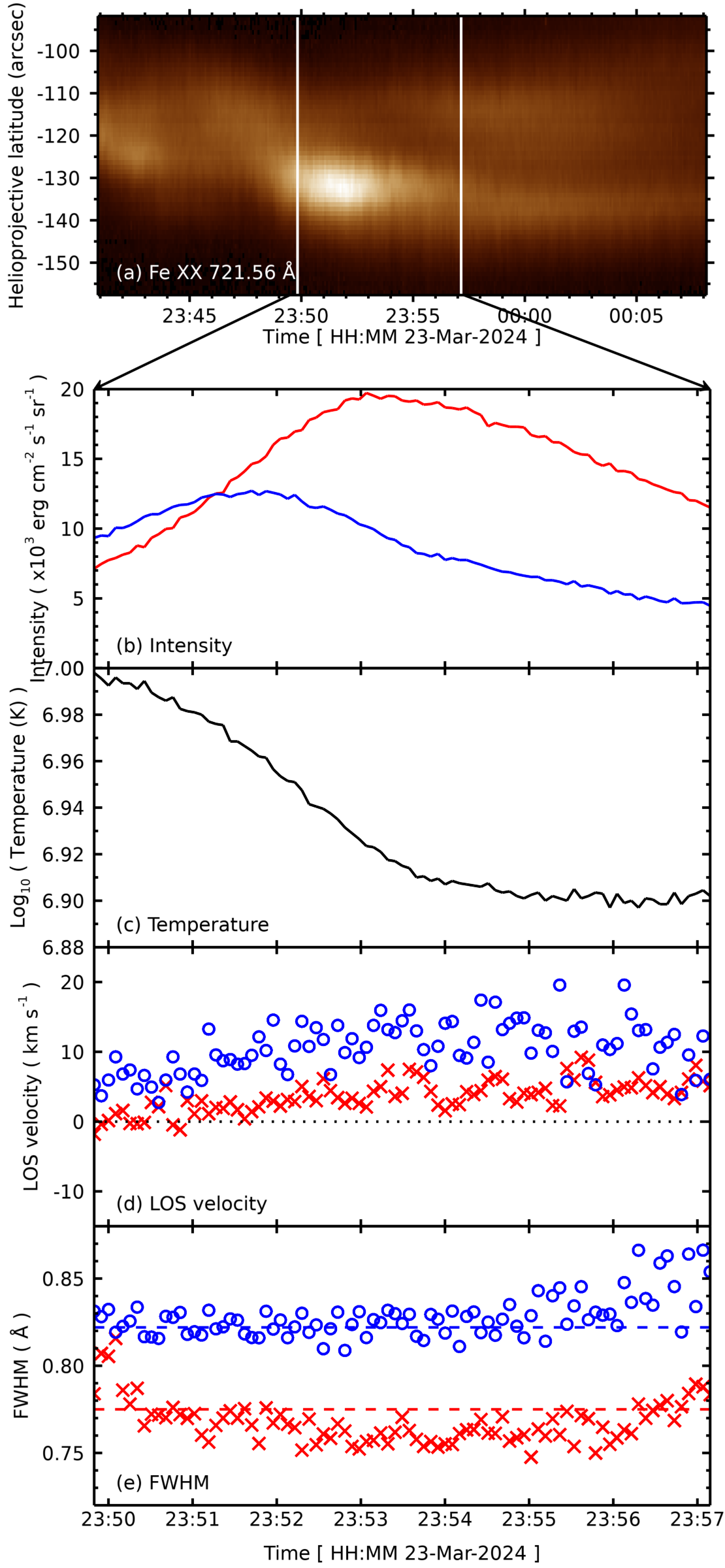}
    \caption{Panel (a) shows the intensity image from \ion{Fe}{xx} 721.56~\AA\ obtained from five consecutive SPICE rasters. Panel (b) shows the variation of intensity with time for the \ion{Fe}{xx} (blue) and \ion{Fe}{xviii} 974.86~\AA\ (red) lines. Panel (c) shows the variation of temperature with time, determined from the \ion{Fe}{xx} / \ion{Fe}{xviii} ratio. Panels (d) and (e) show the variations of LOS velocity and FWHM for the two lines. The FWHM values have been corrected for thermal broadening based on the temperatures shown in Panel (c). The dashed lines on Panel (e) show the instrumental widths for the SW (blue) and LW (red) channels (Appendix~\ref{app:calib}). }
    \label{fig:params}
\end{figure}

Figure~\ref{fig:params} plots various parameters derived from the fits to the \ion{Fe}{xviii} and \ion{Fe}{xx} lines. Panel (a) a shows the \ion{Fe}{xx} intensity image obtained from five consecutive rasters, beginning at 23:36~UT. The vertical lines highlight a time region where \ion{Fe}{xx} is brightest and the two hot line intensities peak at the same $y$-pixel. Panels (b), (d) and  (e) show the variation of the narrow Gaussian fit parameters over this time region for \ion{Fe}{xviii} and \ion{Fe}{xx}, shown in red and blue, respectively. The ratio of the two line intensities shown in Panel (b) is converted to temperature using the curve shown in Figure~\ref{fig:chianti}. The temperature is seen to fall by 0.08~dex over a five minute period, before plateauing at 23:54~UT. Based on estimates of the flare loop size and density (Appendix~\ref{app:cooling}), we estimate thermal conduction and radiative cooling times of 170~s and 250~s, respectively, for the loop to cool from 10~MK to 8~MK. As the actual cooling time is significantly longer than the conductive timescale, then it suggests continued heating is responsible for slowing down the cooling time.

LOS velocities for the iron lines were derived using  the reference wavelengths given in Section~\ref{sec:blend}, and large differences of 30--40~\kms\ between the velocities of the two ions were found. This is not expected for two lines that are relatively close in temperature, and Appendix~\ref{app:velcalib} shows that a similar discrepancy is found for \ion{O}{vi} 1031.9~\AA\ and \ion{Ne}{viii} 770.4~\AA, two strong transition region lines in the SW and LW channels. The latter pair of lines are used to determine a correction to the SPICE wavelength scale, and the \ion{Ne}{viii} line is used as the reference in setting the zero velocity scale. Figure~\ref{fig:params}(d) shows the resulting LOS velocities for \ion{Fe}{xviii} and \ion{Fe}{xx}. The former ion shows velocities close to zero, with a small increase in time. The uncertainty in the velocity measurements is $<2$~\kms, demonstrating that SPICE is capable of precise LOS velocity measurements despite the modest spectral resolution of the instrument and short exposure time of this observation. The \ion{Fe}{xx} line is red-shifted relative to \ion{Fe}{xviii} by, on average, 7~\kms\ although, given the uncertainty in setting the velocity scale (Appendix~\ref{app:velcalib}), this is probably not significant. The scatter in the \ion{Fe}{xx} velocities is around 3.5~\kms, which is larger than for \ion{Fe}{xviii} and also larger than the statistical uncertainties of 1.0--1.5~\kms, which suggests an additional source of uncertainty not accounted for in the SPICE calibration pipeline.

The FWHM values of the narrow Gaussian components from the line fits are corrected for thermal broadening by subtracting in quadrature the thermal widths calculated for the temperatures obtained from the line ratio diagnostic (Figure~\ref{fig:params}(c)). The resulting line widths are plotted in Figure~\ref{fig:params}(e) where they can be compared with the instrumental widths that were estimated from transition region ions as described in Appendix~\ref{app:calib}. The \ion{Fe}{xx} line width is very close to the instrumental width except in the last two minutes of the time sequence where it increases by 20--30~m\AA. The \ion{Fe}{xviii} line width is mostly below the instrumental width by 10--20~m\AA. This may indicate a change in the LW instrumental width since 2020, or it may be a consequence of fitting three emission lines simultaneously for this feature leading to a systematic underestimate of the line width. \citet{2015ApJ...799..218Y} found non-thermal velocities of 30--40~\kms\ for the \ion{Fe}{xxi} 1354~\AA\ line observed by IRIS in flare loops. Adding a 40~\kms\ non-thermal velocity to the SPICE lines would increase the FWHM values above the instrumental widths by 15~m\AA\ and 30~m\AA\ for \ion{Fe}{xx} and \ion{Fe}{xviii}, respectively. The SPICE line width measurements are sufficiently precise to rule this out for \ion{Fe}{xviii}, except perhaps at the beginning and end of the time interval. However, further work on characterizing the SPICE instrumental line width and how it changes with time will be required before definitive conclusions on non-thermal velocities from the hot iron lines can be made.

\section{Summary and Conclusions}\label{sec:summary}

The \ion{Fe}{xviii} 974.86~\AA\ and \ion{Fe}{xx} 721.56~\AA\ forbidden lines have been measured together in a solar flare observed by SPICE for the first time. The lines were fit with functions comprising a narrow and broad Gaussians that were derived empirically from a flare spectral atlas dataset. The ratio of the two lines was used to produce a map showing the variation of temperature along the SPICE slit as a function of time (Figure~\ref{fig:temp}). This revealed episodic heating and motions in the $y$-direction that are consistent with the apparent untwisting of loops seen in AIA images (Figure~\ref{fig:movie}).  A seven minute period near the peak brightnesses of the two lines was selected, and line fit parameters derived (Figure~\ref{fig:params}). The intensity ratio of the two lines was converted to an isothermal plasma temperature, revealing a decreasing temperature from 10~MK to 8~MK in a five minute interval, which is slower than expected for radiative and conductive cooling, suggesting continued heating during the cooling phase.  Doppler velocities in the range 0 to $+10$~\kms\ were derived, but uncertainties in the absolute wavelength calibration prevent us from concluding if there are plasma flows in the loop. The widths of the two lines are very close to the instrumental widths for both lines during the period when the lines are brightest, and there is no evidence for non-thermal broadening although further work on characterizing the SPICE instrumental width is required to confirm this.

Both lines yielded strong signal in the 5~s exposures used for the observation and demonstrate that $\sim$~10~MK plasma can be observed at high cadence with SPICE, potentially as low as 1~s. The \ion{Fe}{xx} line is partially blended with \ion{Fe}{viii} 721.27~\AA, but is found to be negligible here. The blend may be more significant in flare ribbons, where transition region lines are strongly enhanced.

The Solar-C EUV Solar Telescope \citep{2020SPIE11444E..0NS}, to be launched in 2028, will observe the \ion{Fe}{xviii} and \ion{Fe}{xx} lines at spectral and spatial resolutions an order of magnitude higher than SPICE, and with an effective area around a factor 100 larger. The present study has demonstrated that the lines will be excellent flare diagnostics for Solar-C.

\begin{acknowledgements}
The anonymous referee is thanked for valuable comments and suggestions. The authors acknowledge NASA funding for the SPICE instrument team at GSFC. GSK acknowledges support from a NASA Early Career Investigator Program award (grant \# 80NSSC21K0460). 
The development of SPICE has been funded by ESA member states and ESA. It was built and is operated by a multi-national consortium of research institutes supported by their respective funding agencies: STFC RAL (UKSA, hardware lead), IAS (CNES, operations lead), GSFC (NASA), MPS (DLR), PMOD/WRC (Swiss Space Office), SwRI (NASA), UiO (Norwegian Space Agency).
\end{acknowledgements}

\facilities{Solar Orbiter(SPICE), SDO(AIA)}

\bibliography{ms}{}
\bibliographystyle{aasjournal}

\appendix

\section{Velocity calibration}\label{app:velcalib}

The  Gaussian line fit parameters derived for the \ion{Fe}{xviii} and \ion{Fe}{xx} lines (Section~\ref{sec:gauss}) yielded Doppler velocities that were different by around 30--40~\kms, with the \ion{Fe}{xx} line redshifted relative to \ion{Fe}{xviii}. Such a large difference is not expected for two lines relatively close in temperature, and it is significantly larger than the 3--4~\kms\ uncertainty on the lines' reference wavelengths (Section~\ref{sec:blend}). Therefore, it is likely there is a problem with the wavelength scale of the SPICE instrument, and in particular there may be a wavelength offset between the SW and LW channels.

\begin{figure}[t]
    \centering
    \includegraphics[width=0.7\linewidth]{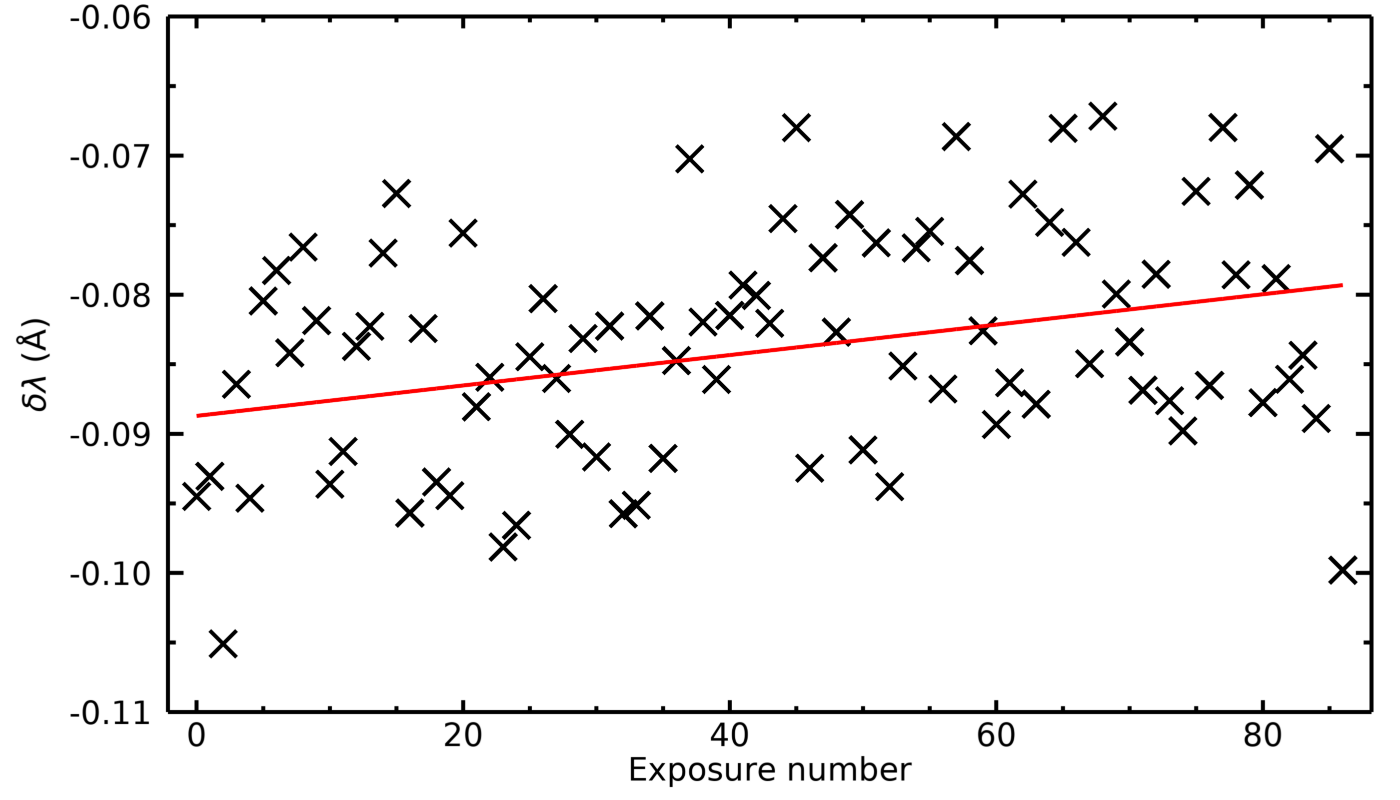}
    \caption{Plot of the \ion{O}{vi}--\ion{Ne}{viii} wavelength offset, $\delta\lambda$, as a function of exposure number for the time period corresponding to that shown in Panels (b)--(e) of Figure~\ref{fig:params}. The red line is a linear fit to the data.}
    \label{fig:o6-ne8}
\end{figure}

To investigate this, the strong \ion{Ne}{viii} 770.4~\AA\ and \ion{O}{vi} 1031.9~\AA\ lines were considered. The lines lie in the SW and LW channels, respectively, and were observed in windows 4 and 8 of the flare sit-and-stare study. For the reduced time period highlighted in Figure~\ref{fig:params}, the \ion{Ne}{viii} and \ion{O}{vi} lines were fit at each pixel along the SPICE slit using the line profile models described in Section~\ref{sec:profile}. For each line and each exposure, the average centroid along the slit was calculated, rejecting pixels where one or both of the line fits had reduced $\chi^2$ values greater than 1.5. Writing the average centroids as $\lambda_i^\mathrm{SW}$ and $\lambda_i^\mathrm{LW}$, for \ion{Ne}{viii} and \ion{O}{vi}, respectively, and exposure number $i$, we consider the wavelength offset $\delta\lambda = (\lambda_i^\mathrm{LW} -\lambda_\mathrm{ref}^\mathrm{LW})-(\lambda_i^\mathrm{SW} -\lambda_\mathrm{ref}^\mathrm{SW})$, where the $\lambda_\mathrm{ref}$ are the reference wavelengths for the two lines. For the latter we used the values 770.429~\AA\ and 1031.927~\AA\ from the NIST database \citep{NIST_ASD}. Figure~\ref{fig:o6-ne8} plots $\delta\lambda$ for the 87 SPICE exposures. The mean value is $-83.5$~m\AA, and there is evidence of an increase with time as shown by the linear fit shown on Figure~\ref{fig:o6-ne8}. The offset implies that the \ion{Ne}{viii} line is systematically red-shifted relative to the \ion{O}{vi} line by 32~\kms. This offset is consistent with that found for the iron lines.

As the lines are formed relatively close in temperature, with formation temperatures of 0.6~MK and 0.3~MK respectively, this offset is unexpected and suggests a systematic offset in the SPICE wavelength calibration scale. The slope of the linear fit implies a slight drift in the offset from 35~\kms\ to 31~\kms\ but the scatter in the measurements is quite large, at 22~m\AA\ or 9~\kms.

The mean value of $(\lambda_i^\mathrm{LW} -\lambda_\mathrm{ref}^\mathrm{LW})$  across the 87 exposures is $-80$~m\AA\ or $-23$~\kms, and the mean value of  $(\lambda_i^\mathrm{SW} -\lambda_\mathrm{ref}^\mathrm{SW})$ is $+3.6$~m\AA, or $+1.4$~\kms. We may expect the average velocity over the raster to be close to zero, and so we infer that the SW channel wavelength calibration is closer to the absolute reference frame and hence that the LW channel wavelength should be adjusted.

For the \ion{Fe}{xviii} and \ion{Fe}{xx} lines, the \ion{Fe}{xx} measured centroids are retained while the \ion{Fe}{xviii} wavelengths are adjusted on an exposure-by-exposure basis using the straight line fit shown on Figure~\ref{fig:o6-ne8}. The resulting Doppler velocities are shown in Figure~\ref{fig:params}(d).

\section{Instrumental line width}\label{app:calib}

Instrumental widths for the SPICE 2\as\ slit were given by \citet{2021A&A...656A..38F}, but these were found to be inconsistent with the line widths in the present work from data obtained with the 4\as\ slit. In particular, the \citet{2021A&A...656A..38F} widths correspond to 0.761~\AA\ and 0.904~\AA\ for the SW and LW channels, respectively, which are significantly smaller and larger, respectively, than the widths for the flare dataset. In this Appendix we estimate instrumental widths for the two channels using a quiet Sun dataset obtained on 28 May 2020 at 16:05~UT. This was one of the earliest rasters obtained by SPICE and was chosen to minimize the effects of detector burn-in. The latter is a depression in the sensitivity of the detector caused by prolonged solar exposure that results in a flat-top to the line profiles since the effect is strongest at the center of the emission line. If uncorrected the measured line width will be broader than the expected line width. The SPICE calibration pipeline corrects for burn-in. However, as emission lines move on the SPICE detector during the spacecraft orbit, the burn-in correction likely does not fully retrieve the original line profile.

\begin{figure}[t]
    \centering
    \includegraphics[width=0.7\linewidth]{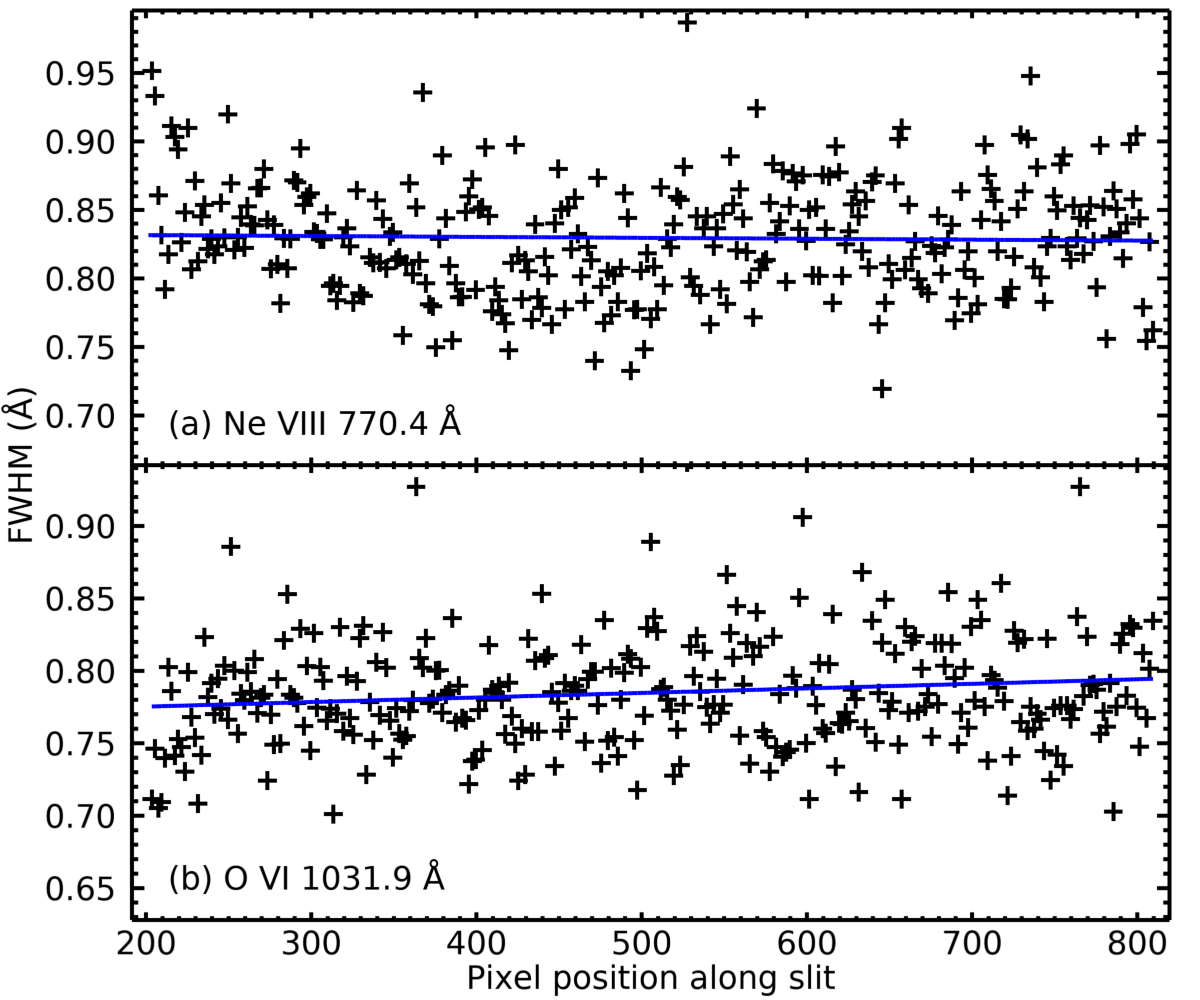}
    \caption{Plots of the  (a) \ion{Ne}{viii} 770.4~\AA\ and (b) \ion{O}{vi} 1031.9~\AA\ FWHM values as a function of the SPICE detector $y$ pixel for the 2020 May 28 quiet Sun dataset. Each cross represents the median value of the FWHM in the raster $x$ direction. The blue lines are the linear fits to the FWHM values.} 
    \label{fig:widths}
\end{figure}

The 28 May 2020 dataset is the same one studied by \citet{2021A&A...656A..38F}, where further details can be found. The wavelength windows for the strong \ion{Ne}{viii} 770.4~\AA\ and \ion{O}{vi} 1031.9~\AA\ lines were extracted and rebinned spatially by 8 pixels in $x$ and 2 pixels in $y$. The broadened Gaussian fit functions described in Section~\ref{sec:profile} were fit to the lines at each macropixel. The median FWHM value along each row of the image was obtained to yield FWHM values as a function of $y$, and straight lines were fit to the two FWHM distributions. The results are shown in Figure~\ref{fig:widths}. The \ion{Ne}{viii} line widths showed no significant slope, whereas the \ion{O}{vi} widths showed a small slope of 0.031~m\AA~pix$^{-1}$ (an increase of 19~m\AA\ from $y$ pixel 200 to $y$ pixel 800). The mean widths along the slit are 0.830~\AA\ and 0.785~\AA\ for \ion{Ne}{viii} and \ion{O}{vi}, respectively. The widths are much larger than the thermal and non-thermal widths of these lines, but we correct for these to yield final instrumental width values of 0.822~\AA\ and 0.775~\AA, respectively. The thermal widths are obtained assuming the ions are formed at the temperature of maximum ionization given by CHIANTI. For the non-thermal widths we used the disk center value of 11~\kms\ measured by \citet{1999ApJ...516..490P} from SUMER measurements of \ion{Ne}{viii} 770.4~\AA. 
The final instrumental widths are over-plotted on Figure~\ref{fig:params}(e) as dashed lines.

\section{Loop cooling times}\label{app:cooling}

In Figure~\ref{fig:params}(c) the flare loop temperature is seen to drop from 10~MK to 8~MK in five minutes. The dominant cooling processes for a flare loop are thermal conduction and radiative losses, and simple formulae for estimating their contributions are given in \citet{1995ApJ...439.1034C}. To apply the formulae it is necessary to have estimates of the loop half length, $L$, and  the electron density, $n$. $L$ is estimated from the AIA 94~\AA\ image at 23:51~UT (Figure~\ref{fig:movie}) where the ends of the bright loop are separated by 61\as. For a semi-circular loop shape we then have $L=35$~Mm.

Based on the intensity cross-section of the flare loop in the SPICE \ion{Fe}{xx} emission line and the AIA 94~\AA\ image, the diameter of the loop is estimated to be 4~Mm. If the loop has a circular cross-section, then the plasma column depth, $h$, at the center of the loop is the same as the diameter. The  electron density can then be estimated from the formula
\begin{equation}
     I = 0.85 \varepsilon G n^2 h
\end{equation}
where $I$ is the \ion{Fe}{xx} line intensity at the loop center, the factor 0.85 is the ratio of protons to electrons in the corona, $\varepsilon$ is the abundance of iron relative to hydrogen, and $G$ is the value of the \ion{Fe}{xx} contribution function at 10~MK. $I=1.15\times 10^4$~\ecss, $\varepsilon=2.88\times 10^{-5}$ using the photospheric abundances of \citet{2021A&A...653A.141A}, and $G=8.38\times 10^{-23}$~erg\,cm$^{3}$\,s$^{-1}$ from CHIANTI. We therefore have $n=1.2\times 10^{11}$~cm$^{-3}$, and the thermal conduction cooling timescale, $\tau_\mathrm{c}$, is 1830~s using Equation 2 of \citet{1995ApJ...439.1034C}. From Equation~3a of \citet{1995ApJ...439.1034C}, the time to cool from 10~MK to 8~MK is 170~s. 

The radiative loss function, $Q(T)$, was derived using the CHIANTI \textsf{rad\_loss} IDL function, and is approximately constant between 8~MK and 10~MK with a value of $2.8\times 10^{-23}$~erg~cm$^3$~s$^{-1}$. The radiative loss cooling time, $\tau_\mathrm{r}=3kT/(0.85 nQ(T))$, where $k$ is Boltzmann's constant, is 1220~s. Using Equation 4 of \citet{1995ApJ...439.1034C}, the time taken to cool from 10~MK to 8~MK is then 250~s. 

Radiative cooling and thermal conduction are therefore both expected to play a role in the cooling of the loop, with conduction dominating. If coronal abundances are assumed instead of photospheric abundances, then the conductive cooling time would reduce by around a factor two and the radiative cooling time would increase by a similar factor.

\end{document}